\input article.sty\relax
\@addtoreset{equation}{section}
\newcounter{subeqn}

\documentstyle[12pt,dina4]{myart}

\newcommand{\be}{\begin{equation}}
\newcommand{\ee}{\end{equation}}
\newcommand{\bea}{\begin{eqnarray}}
\newcommand{\eea}{\end{eqnarray}}
\newcommand{\bfr}{\begin{flushright}}
\newcommand{\efr}{\end{flushright}}

\newcounter{subeq}

\newcommand{\subeqno}{\stepcounter{subeq} \addtocounter{equation}{-1}}

\newcommand{\subeqres}{\setcounter{subeq}{0}}
\newcommand{\eqnoinc}{\addtocounter{equation}{1}}
\begin{document}

\title
{
Vibrations versus collisions and the iterative structure of
two-body dynamics\footnote{supported by DFG and GSI Darmstadt}
}
\author{A. Pfitzner$^1$, W. Cassing$^2$ and A. Peter$^2$ \\
$^1$Forschungszentrum Rossendorf \\ Institut f\"{u}r Kern- und Hadronenphysik,
 D-01314 Dresden \\
 $^2$Universit\"{a}t Giessen, Institut f\"{u}r Theoretische Physik \\
 D-35392 Giessen}
\maketitle
\begin{abstract}
We adopt a truncated version of two-body dynamics by neglecting three-body
correlations, as is supported by microscopic numerical calculations.
Introducing orthogonal channel correlations
for the $pp$- and the $ph$-channel and integrating the latter in terms of
vibrational RPA-states we derive a retarded two-body equation. Its solution
is nonperturbative with respect to loops, ladders and mixed contributions.
In the stationary limit we obtain
an equation for a generalised effective interaction which iterates both
the $G$-matrix and the polarisation matrix.
An in-medium scattering approach transparently demonstrates the
collisional damping of the vibrations.
\end{abstract}
\newpage
\section{Introduction}
A still outstanding problem in
two-body dynamics concerns the simultaneous nonperturbative consideration
of $pp$- ($hh$-) and $ph$-interactions, i.e. the consistent integration
of both the $pp$- and the $ph$-channel.
As numerically tested in \cite{1} within the concept
of correlation dynamics (\cite{2}-\cite{15a}) the short-range part of
the two-body interaction produces correlations mainly in the $pp$-channel,
while the long-range part favours the $ph$-channel. These results are in
agreement with earlier discussions by Abrikosov et al. \cite{5} for infinite
nuclear matter and by Migdal \cite{6} for finite nuclei which show that the sum
of
ladder diagrams should be essential at large momentum transfer whereas one
expects loop diagrams to be dominant at small momentum transfer. In this
sense, the simultaneous consideration of both channels corresponds to a
consistent treatment of short- and long-range correlations (including their
mixing), and therefore of collisional and collective aspects in nuclear
dynamics. \\
Since a numerical solution of the equation of motion for the two-body
correlations is available so far only for light nuclei and in a restricted
single-particle basis it is of current interest to make the iterative
structure of this equation more transparent, in particular with respect to
the mutual influence of short- and long-range correlations which is
not included in $G$-matrix \cite{1a}-\cite{6a} or RPA calculations
\cite{10a}-\cite{14a} which specialize on short-range or
long-range correlations, respectively.
This should
help for a better understanding of the interplay between collisions and
vibrations during the evolution. Formally, the mutual influence shows up
in mixed diagrams combining ladders and loops, a problem that
has been also addressed within the framework of
Green's functions \cite{16a,17a}. The importance for the evaluation
of occupation number distributions and for the damping width of
single-particle and collective excitations
(especially for finite temperatures as e.g. investigated
experimentally in (c.f. \cite{1e,2e}) and theoretically in
(c.f. \cite{1d}-\cite{3d})) has been  demonstrated in
\cite{1} within a model appropriate for light nuclei as well as within the
microscopic TDDM approach \cite{DeBlasio}.

The paper is organised as follows: In section 2 we present microscopic
studies on the relative importance of 3- and 4-body
correlations and on the effect of channel mixing for the total
energy in case of a light model system ($^{16}O$). In section 3 coupled
equations for channel correlations are derived and their iterative structure
is made transparent by introducing vibrational RPA-states. Section 4 is
devoted to the stationary limit while section 5 investigates collisional
damping within an in-medium scattering approach.
The paper is summarised in section 6.
%
%
\section{Microscopic numerical studies}
The nuclear many-body problem on the two-body level can be formulated in
terms of coupled equations of motion for the one-body density $\rho$ and
the two-body correlation function $C$ which read in an arbitrary
single-particle basis \cite{1,3}:
\bea
i\frac{\partial}{\partial t} \rho_{\alpha\alpha'} =
\langle\alpha|[h(1),\rho]|\alpha'\rangle +
\sum_{\beta} \langle\alpha\beta|[v,{\hat C}]|\alpha'\beta\rangle ,
\label{2.1}
\eea

\bea
i\frac{\partial}{\partial t} C_{\alpha\beta\alpha'\beta'} =
\langle\alpha\beta|[h(1)+h(2),{\hat C}]|\alpha'\beta'\rangle
\nonumber\\ \label{2.2} \eqnoinc
+ \langle\alpha\beta|[v^=,\rho_{20}]|\alpha'\beta'\rangle \qquad\qquad
\subeqno \label{2.2a}
\\
+ \langle\alpha\beta|[v^=,{\hat C}]|\alpha'\beta'\rangle  \qquad\qquad
\subeqno \label{2.2b}
\\
+ {\cal A}_{\alpha\beta} {\cal A}_{\alpha'\beta'} \sum_{\gamma\gamma'}
v^{\perp}_{\alpha\gamma'\alpha'\gamma} C_{\gamma\beta\gamma'\beta'}
\subeqno \label{2.2c}
\\
+ \sum_{\gamma \gamma'} \delta_{\gamma \gamma'}
({\cal P}_{\alpha\gamma} {\cal P}_{\alpha'\gamma'}
+{\cal P}_{\beta\gamma} {\cal P}_{\beta'\gamma'})  \nonumber \\
 \times \sum_{\lambda \delta}
\{ v_{\alpha\beta \lambda \delta}  \
C^3_{\lambda \delta \gamma \alpha'\beta'\gamma'}
- C^3_{\alpha \beta \gamma \lambda \delta \gamma'}\
v_{\lambda \delta \alpha' \beta'} \} ,
\subeqno \label{2.2d}
\eea
\subeqres
or in short-hand notation (dropping the term (2.2d) in view of section 3)
\bea
i\dot \rho = [h(1),\rho]+tr_{(2)}[v,C]
\label{2.3}
\eea
\bea
i\dot{C} & = & [h(1)+h(2),C]+[v^{=},\rho_{20}+C]+{\cal A}(1+{\cal
P}{\cal P})v^{\perp}C.
\label{2.4}
\eea
Here we have used the bare mean field $h= t+tr(v^{a}\rho)$, where $
t$ is the kinetic energy operator, while $v^a$ is the antisymmetrised
two-body interaction. Further, $\rho_{20}=\cal A\rho\rho$ is the uncorrelated
two-body density, and $\cal A$ and $\cal P$ denote the antisymmetrisation
and permutation operator, respectively (c.f. \cite{1}). The in-medium
interactions
$v^{=}=Q^{=}v$ and $v^{\perp}=Q^{\perp}v^{a}$ are density-dependent via the
blocking operators which read in an arbitrary single-particle basis
$Q^{=}_{\alpha\beta\alpha'\beta'}=\delta_{\alpha\alpha'}\delta_{\beta\beta'}
-\delta_{\alpha\alpha'}\rho_{\beta\beta'}-\delta_{\beta\beta'}\rho_{\alpha
\alpha'}$ and $Q^{\perp}_{\alpha\beta\alpha'\beta'}=\delta_{\alpha\beta'}
\rho_{\beta\alpha'}-\delta_{\beta\alpha'}\rho_{\alpha\beta'}$ (c.f. \cite{7}).
In
a basis that diagonalises $\rho$ one verifies that $Q^{=}$ projects on $pp$-
and $hh$-states in the Hartree-Fock-limit while $Q^{\perp}$ projects
on $ph$- and $hp$-states. This leads to the definition of two ''orthogonal``
channels: dropping the term with $v^\perp C$ in (\ref{2.2}) constitutes
the $pp$ $(hh)$-channel, while dropping of $v^=C$ constitutes
the $ph$-channel.
Hence an inclusion of (\ref{2.2a}) and (\ref{2.2b}) corresponds to a
nonperturbative resummation of ladder diagrams
({\bf TDGMT}\footnote{{\bf T}ime-{\bf D}ependent-{\bf G}-{\bf M}atrix-{\bf
T}heory})
while (\ref{2.2a}) and (\ref{2.2c}) leads to a resummation of loop diagrams
({\bf RPA}\footnote{{\bf R}andom-{\bf P}hase-{\bf A}pproximation}).
The addressed channel mixing is then obtained by taking into account
(\ref{2.2a}$-$\ref{2.2c})
({\bf NQCD}\footnote{{\bf N}uclear-{\bf Q}uantum-{\bf
C}orrelation-{\bf D}ynamics}). The terms (\ref{2.2d}) include the coupling
to the three-body correlations.  An explicit expression for $C_3$ as a function
of $\rho, \rho_2 $ and $C_2$ is given in \cite{1,4} and corresponds to the
lowest
order three-body correlations which are needed to restore the
fundamental trace relations dynamically for spin symmetric systems
(c.f. \cite{1}). The respective
limit (including (2.2a)-(2.2d) is denoted by
{\bf SCD}\footnote{{\bf S}elfconsistent-{\bf C}orrelation- {\bf D}ynamics}.

The calculations in the following subsections are carried out to support
the analytical considerations of this paper and are performed within
a finite oscillator basis as explained in detail in \cite{1}.
\subsection{Importance of higher order correlations}
According to the hierarchy of density matrices \cite{2} the two-body
correlations $C=C_2$ are coupled dynamically with three-body
correlations $C_3$, and so forth. Any truncation thus becomes only
meaningful if the effect of higher order correlations decreases with
increasing rank $n$. In \cite{1} it has been shown in the framework of model
calculations for light nuclei that the dynamical influence of the
trace conserving terms (2.2d) in fact is quite small.
To confirm this result and to give additional support for a
neglect of $C_3$ for nuclear configurations close to the ground state
in the following sections we compare the
trace-relations between different levels of the hierarchy with each other.
For this purpose we calculate the total traces of $C=C_2$, $C_3$ and
$C_4$, respectively, starting from the trace relations for the density
matrices $\rho_{n}$ [2]
\begin{equation}
\rho _n = \frac{1}{(A-n)!} \  tr_{(n+1, \dots, A)} \rho _A =
\frac{1}{(A-n)} \  tr_{(n+1)} \rho _{n+1}.
\label{2.5}
\end{equation}
Within the cluster expansion for $\rho_{n}$ [2] (c.f. Appendix A)
this leads to the equivalent trace relations for the correlation functions
\begin{equation}
tr_{(2=2')} C_2(12,1'2') = - tr_{(2=2')}
(1/A-{\cal P}_{12})\rho (22')\rho (11') ,
\label{2.6}
\end{equation}
\begin{equation}
tr_{(3=3')} C_3(123,1'2'3') =
 -tr_{(3=3')} (2/A -{\cal P}_{13}-{\cal P}_{23}-{\cal P}_{1'3'}-{\cal
P}_{2'3'})\rho (33')C_2(12,1'2'),
\label{2.7}
\end{equation}
\begin{eqnarray}
\label{2.8}
\lefteqn{tr_{(4=4')} C_4(1234,1'2'3'4') = }   \nonumber \\
&& -tr_{(4=4')} (3/A -{\cal P}_{14}-{\cal P}_{24}-{\cal P}_{34}
-{\cal P}_{1'4'}-{\cal P}_{2'4'}-{\cal P}_{3'4'})
\nonumber \\
&& \times \rho (44') C_3(123,1'2'3') \nonumber \\
&& +tr_{(4=4')} ({\cal P}_{14}+{\cal P}_{24}+{\cal P}_{1'4'}+{\cal P}_{2'4'})
C_2(34,3'4') C_2(12,1'2') \nonumber \\
&& +tr_{(4=4')} ({\cal P}_{34}+{\cal P}_{3'4'}) C_2(24,2'4') C_2(13,1'3')
\nonumber \\
&& -tr_{(4=4')} (C_2(12,3'4') C_2(34,1'2') + C_2(23,1'4') C_2(14,2'3')
\nonumber \\
&& + C_2(24,1'3') C_2(13,2'4')).
\end{eqnarray}
Using (\ref{2.6})-(\ref{2.8}) the total traces over all particle coordinates
are evaluated within a single-particle basis that
diagonalises the one-body density, i.e.
\begin{equation}
tr \ C_2 = - \sum_{\alpha} n_{\alpha} (1 - n_{\alpha}),
\label{2.9}
\end{equation}
\begin{equation}
tr \ C_3 = 2 \sum_{\alpha} n_{\alpha} (1 - n_{\alpha})
(1 - 2 n_{\alpha}),
\label{2.10}
\end{equation}
\begin{eqnarray}
\label{2.11}
\lefteqn{tr \ C_4 = -6 \sum_{\alpha} n_{\alpha} (1 - n_{\alpha})
(1 - 2 n_{\alpha}) + 12 \sum_{\alpha} n_{\alpha}^2 (1 - n_{\alpha})^2
+ 12 \sum_{\alpha \beta} n_{\alpha} n_{\beta} C_{\alpha \beta
\alpha \beta} }  \nonumber \\
&& + \sum_{\alpha \beta \gamma \lambda} \{C_{\alpha \beta \alpha \lambda}
C_{\gamma \lambda \gamma \beta} +
C_{\alpha \beta  \lambda \beta} C_{\gamma \lambda \gamma \alpha}
 + C_{\gamma \beta \gamma \lambda} C_{\alpha \lambda \alpha \beta}
+ C_{\lambda \beta \alpha \beta} C_{\gamma \alpha \gamma \lambda} \nonumber \\
&& + C_{\gamma \beta \lambda \beta} C_{\alpha \lambda \alpha \gamma}
+ C_{\lambda \beta \gamma \beta} C_{\alpha \gamma \alpha \lambda}
- C_{\alpha \beta \gamma \lambda} C_{\gamma \lambda \alpha \beta}
- C_{\gamma \beta \alpha \lambda} C_{\alpha \lambda \gamma \beta}
- C_{\beta \lambda \alpha \gamma} C_{\alpha \gamma \beta \lambda}\}.
\nonumber\\
\end{eqnarray}
The time-averaged numerical results for $tr \ C_2$, $tr \ C_3$
 and $tr \ C_4$ are
displayed in Fig. 1 for the $^{16}O$ model system \cite{1}
as a function of the initialization temperature T
which via the Fermi distribution
$\rho_{\alpha\alpha}(t=0)=[1+exp((\epsilon_\alpha - \epsilon_F)/T)]^{-1}$
is directly related to the initial thermal excitation energy.
We observe that the effect of higher-order correlations decreases
by about one order of magnitude in each case. This can be viewed as
a further indication
for the convergence of the cluster expansion within our model and
supports the neglect of $C_3$ in the general considerations of the
following sections. We note, however, that although $tr C_3 $ and
$tr C_4$ might remain small,
individual matrix elements may become quite large and be responsible
for new physical phenomena.
\subsection{Effect of channel mixing on the corrrelated ground state}
A central problem within correlation dynamics is the precise determination
of the correlated ground state \cite{15a}. We address this problem
numerically and approach the correlated ground state
energy by switching on the two-body interactions adiabatically in time.
In order to define a proper adiabatic regime the following  calculations are
performed:
we initialize a system of 16 nucleons in its
Hartree-Fock-configuration at $T=0$ MeV temperature.
Then we switch on the two-body interaction in the term $vC$ in (\ref{2.1})
and the terms $v^{=}C$ and $v^{\perp}C$ in (\ref{2.2}) by multiplying the
interaction with a common function $f(t)=t/t_{s}$ for
$0 \le t \le t_{s}$ or $1$ for $t>t_{s}$, respectively.
Since for a fixed (time independent) two-body interaction the equations
of motion (2.1) and (2.2) conserve the total energy for $t \ge t_s$ the
ground state energy is extracted by evaluating $<E> = tr_{(1)} t \rho
+ 1/2 tr_{(1,2)} v \rho_2$. This adiabatic approach is expected to work
well for approximately spherical configurations as in
case of $^{16}O$ but might become questionable for nuclei with large
density fluctuations (shape coexistence of oblate and prolate
configurations). Furthermore, due to the neglect of four-body correlations
a mixing with alpha-cluster configurations is not expected to occur, too.

In Fig. 2 we show the resulting total asymptotic energy for $t>t_{s}$
as a function of the parameter $t_{s}$ for the limits BORN, TDGMT, RPA,
NQCD and SCD which specifies the adiabatic regime by the
condition $t_{s}\ge 10^{-21}s$ and thus guaranties the closest approach to
the ground state energy of the correlated system in the respective limit.
It becomes obvious that neither the $pp$-channel
alone (TDGMT) nor the $ph$-channel alone (RPA) is sufficient to come
close to the lowest possible energy which corresponds to the full theory
(SCD). Thus channel-mixing is an important aspect of nuclear dynamics
and demands for a deeper understanding. On the other hand the limits NQCD and
SCD
practically give the same result for the ground state energy such that the
three-body terms in (2.2d) - which garantee the dynamical conservation of
the trace relations - can be neglected for the following investigations.
\section{Vibrational retardation and the iterative structure of channel mixing}
In order to associate with each channel its own correlation function we
subdivide the correlation function $C$ into
channel correlations according to
\be
C = C^{=} + C^{\perp}.
\label{3.1}
\ee
The channel correlations then follow the coupled set of equations
\bea
i\dot{C}^= & = & [H^{=}_{0},C^{=}]+[v^{=},C^{=} + C^{\perp} + \rho_{20}]
\label{3.2}
\eea
\bea
i\dot{C}^{\perp} & = & (1+{\cal P}{\cal P})H^{\perp}_{0}C^{\perp}
+\nonumber\\ & &+{\cal A}(1+{\cal P}{\cal P})v^{\perp}(C^{\perp }+C^{=})\; ,
\label{3.3}
\eea
or, in an arbitrary single-particle basis,
\bea
i\dot{C}^{=}_{\alpha\beta\alpha{'}\beta{'}}&=&
     \langle\alpha\beta|[H^{=}_{0},C^{=}]|\alpha{'}\beta{'}\rangle
+\langle\alpha\beta|[v^{=},C^{=}+C^{\perp}+\rho_{20}]|\alpha{'}\beta{'}\rangle
\label{3.4}\\ i\dot{C}^{\perp}_{\alpha\beta\alpha{'}\beta{'}}&=&
(1+{\cal P}_{\alpha\beta}{\cal
P}_{\alpha'\beta'})\sum_{\gamma\gamma'}
H^{0\perp}_{\alpha\gamma{'}\alpha{'}\gamma}
C^{\perp}_{\gamma\beta\gamma{'}\beta{'}}+ \nonumber\\ & & +{\cal
A}_{\alpha\beta}(1+{\cal
P}_{\alpha\beta}{\cal P}_{\alpha{'}\beta{'}})\sum_{\gamma\gamma{'}}
v^{\perp}_{\alpha\gamma{'}\alpha{'}\gamma}(C^{\perp}_{\beta\gamma\beta{'}
\gamma{'}} +C^{=}_{\beta\gamma\beta{'}\gamma{'}}).
\label{3.5}
\eea
Observing that the mean-field contributions in (\ref{3.4})
and (\ref{3.5}) with
\bea
\langle\alpha\beta|H^{=}_{0}|\gamma\gamma{'}\rangle =
h_{\alpha\gamma}\delta_{\beta\gamma{'}}+\delta_{\alpha\gamma}h_{\beta\gamma{'}}
\label{3.6}
\eea
\bea
H^{0\perp}_{\alpha\gamma'\alpha'\gamma}
=
h_{\alpha\gamma}\delta_{\gamma{'}\alpha{'}}-\delta_{\alpha\gamma}h_{\gamma{'}\al
pha{'}}
\label{3.7}
\eea
may be written either in ``horizontal''or ``vertical'' form one easily
proves that the sum of (\ref{3.4}) and (\ref{3.5}) reproduces the original
equation (\ref{2.2}) for $C_{\alpha\beta\alpha{'}\beta{'}}$ .
Eqs. (\ref{3.4}) and (\ref{3.5}) describe the evolution of correlations in
the $pp$- and $ph$-channel, respectively.
Their mutual coupling is accounted for by the inhomogeneous terms which
comprise $C^{\perp}$ in the $pp$- and $C^{=}$ in the $ph$-channel.

The explicite time-dependence of the channel correlations is controlled by
the structure of the two-body equations (\ref{3.4}) and (\ref{3.5}).
In addition we observe an implicite time-dependence due to the one-body
density contained in the mean-field potentials and the in-medium interactions.
To obtain a closer insight into the different time-scales we investigate
numerically, within the same model adopted in section 2, the monopole
response in the occupation numbers and characteristic correlation
matrix elements. From the comparison of Fig. 3 and Fig. 4 it is obvious
that the relative change in time of the occupation numbers is small
compared to that of the two-body matrix elements.
To facilitate the discussion of the iterative structure of the correlations
we thus can disregard the time-dependence of $\rho_{\alpha\alpha'}(t)$
by assuming a stationary
$\rho$ in (\ref{3.4}) and (\ref{3.5}) which means a
decoupling of $C$ from the one-body evolution given
by (\ref{2.1}). This allows to adopt a single-particle basis with
$h_{\alpha\gamma}\approx\delta_{\alpha\gamma}\epsilon_{\alpha}$ and
$\rho_{\alpha\gamma}=\delta_{\alpha\gamma}n_{\alpha}$.
Neglecting the exchange term in the $ph$-channel by dropping
$\cal A_{\alpha\beta}$, eq. (\ref{3.5}) reads
\bea
i\dot{C}^{\perp}=(1+{\cal P}{\cal P})H^{\perp}C^{\perp}+(1+{\cal
P}{\cal P})v^{\perp}C^=,
\label{3.8}
\eea
with the RPA-hamiltonian
$H^{\perp}=H^{\perp}_{0}+v^{\perp}$ and its matrix elements
(in the adopted single-particle basis)
\bea
H^{\perp}_{\alpha\gamma{'}\alpha{'}\gamma}=(\epsilon_{\alpha}-\epsilon_{\alpha{'
}})
\delta_{\alpha\gamma}\delta_{\alpha{'}\gamma{'}}+(n_{\alpha{'}}-n_{\alpha})
v^a_{\alpha\gamma{'}\alpha{'} \gamma}.
\label{3.9}
\eea
For an integration of (\ref{3.8}) it is suggestive to transform this
equation into the RPA-basis.
The RPA-eigenstates $\chi^\mu$ and $\chi^{\mu*}$ (collective phonons) are
given by the secular equations
\bea
\sum_{\gamma\gamma'}H^{\perp}_{\alpha\gamma'\alpha'\gamma}\chi^\mu_{\gamma
\gamma'} &=&\Omega_{\mu}\chi^\mu_{\alpha\alpha'}\; ,
\label{3.10}\\
\sum_{\gamma\gamma'}\chi^{\mu*}_{\gamma\gamma'}H^{\perp\dag}_{\gamma'\beta
\gamma\beta'}
&=&\Omega_\mu\chi^{\mu*}_{\beta\beta'} \; ,
\label{3.11}
\eea
where the matrix elements of the hermitean conjugate operator $H^{\perp\dag}$
are related to those of $H^{\perp}$ by
\be
H^{\perp\dag}_{\gamma'\beta\gamma\beta'}=-H^{\perp}_{\beta\gamma'\beta'\gamma}.
\label{3.12}
\ee
Making use of the orthogonality and completeness relation (c.f. \cite{8})
\bea
\sum_{\alpha\alpha'}\tilde{\chi}^{\mu*}_{\alpha\alpha'}\chi^{\nu}_{\alpha\alpha'
}
& = & \delta_{\mu\nu}N_{\mu}
\label{3.13}\\
\sum_{\mu}N_{\mu}\chi^{\mu}_{\alpha\alpha'}\chi^{\mu*}_{\beta\beta'}
& = & \delta_{\alpha\beta}\delta_{\alpha'\beta'}(n_{\alpha'}-n_{\alpha})\; ,
\label{3.14}
\eea
the relations between matrix elements in the single-particle-  and the
RPA-basis
read
\bea
C^{\perp}_{\alpha\beta\alpha'\beta'}&=&\sum_{\mu\nu}N_{\mu}N_{\nu}\chi^{\mu}_{\a
lpha
\alpha'}\chi^{\nu*}_{\beta\beta'}C_{\mu\nu}
\label{3.15} \\
C_{\mu\nu}&=&\sum_{\alpha\alpha'\beta\beta'}\tilde
{\chi}^{\mu*}_{\alpha\alpha'} \tilde
{\chi}^{\nu}_{\beta\beta'}C_{\alpha\beta\alpha'\beta'}^{\perp}\;,
\label{3.16}
\eea
with
$N_{\mu}=1\;(-1)$ for $\mu>0\;(<0)$ and $\tilde \chi^{\mu}_{\alpha\alpha'}=
(n_{\alpha'}-n_\alpha)^{-1}\chi^\mu_{\alpha\alpha'}$. The equation for
$C_{\mu\nu}$ follows from (\ref{3.16}) using (\ref{3.8})
\bea
i\dot{C}_{\mu\nu}&=&(\Omega_{\mu}-\Omega_{\nu})C_{\mu\nu}+
\sum_{\gamma\gamma'\lambda\lambda'}K^{\mu\nu}_{\gamma\lambda\gamma'\lambda'}
C^{=}_{\gamma\lambda\gamma'\lambda'}
\label{3.17} \\
K^{\mu\nu}_{\gamma\lambda\gamma'\lambda'}&=&\tilde{\chi}^{\nu}_{\lambda
\lambda'}\theta^{\mu*}_{\gamma\gamma'}+\tilde{\chi}^{\mu*}_{\gamma\gamma'}
\theta^{-\nu}_{\lambda\lambda'}\;,
\label{3.18}
\eea
where we have introduced the phonon-particle vertices
\bea
\theta^{\mu}_{\gamma\gamma'}:=\sum_{\alpha\alpha'}v^{a}_{\gamma\alpha'
\gamma'\alpha}\chi^{\mu}_{\alpha\alpha'}&=&(\Omega_{\mu}-\epsilon_{\gamma}
+\epsilon_{\gamma'})\tilde{\chi}^{\mu}_{\gamma\gamma'}
\label{3.19}   \\
\theta^{-\nu}_{\lambda\lambda'}:=\sum_{\beta\beta'}\chi^{\nu}_{\beta\beta'}
v^{a}_{\lambda'\beta\lambda\beta'}&=&(\Omega_{-\nu}-\epsilon_{\lambda}
+\epsilon_{\lambda'})\tilde{\chi}^{\nu}_{\lambda\lambda'}
\label{3.20}
\eea
with $\Omega_{-\nu} = - \Omega_{\nu} $.
These relations  may be proved using the RPA-equations (\ref{3.10}) and
(\ref{3.11}). We note that a conceptionally similar approach has been
given by Belyaev almost three decades ago \cite{Bel1,Bel2} starting from
an expansion of the two-body interaction in terms of spherical tensor
operators.
This leads directly to phonon-particle vertices which induce a nonlocality
in time for the particle propagation due to the phonon exchange diagrams.
Especially the quadrupole-quadrupole term in the two-body interaction is
found to be important for the nuclear structure at low excitation energy
thus also pointing out the importance of mixed diagrams for
nuclear configurations close to the ground state. In this respect
our present analysis
can be regarded as an alternative formulation of the influence of
channel mixing, however, emerges quite naturally within the NQCD equations
without any severe restrictions.

After integration of (\ref{3.17}) the resulting solution $C_{\mu\nu}(t)$
is transformed back into the single-particle basis yielding
($\Omega_{\mu\nu} =\Omega_{\mu}-\Omega_{\nu}$)
\bea
C^{\perp}_{\alpha\beta\alpha'\beta'}(t)&=&\sum_{\mu\nu}N_{\mu}N_{\nu}
\chi^{\mu}_{\alpha\alpha'}\chi^{\nu*}_{\beta\beta'}e^{-i\Omega_{\mu\nu}t}
C_{\mu\nu}(0)+ \nonumber\\
& & +
1/i\int_{0}^{t}dt'\sum_{\mu\nu}N_{\mu}N_{\nu}\chi^{\mu}_{\alpha\alpha'}
\chi^{\nu*}_{\beta\beta'}e^{-i\Omega_{\mu\nu}(t-t')}
\sum_{\gamma\gamma'\lambda\lambda'}K^{\mu\nu}_{\gamma\lambda\gamma'\lambda'}
C^{=}_{\gamma\lambda\gamma'\lambda'}(t').   \nonumber\\
\label{3.21}
\eea
The first term is a superposition of oscillations around the initial values
$C_{\mu\nu}(0)$ and represents a solution of the homogeneous equation that
follows when dropping the term with $C^{=}$ in (\ref{3.8}). The second term
comprises the coupling with the $pp$-channel.

Insertion of (\ref{3.21}) into (\ref{3.4}) leads to a retarded two-body
equation in the $pp$-channel,
\bea
i\dot{C}^{=}_{\alpha\beta\alpha'\beta'}&=&<\alpha\beta|[H^{=},C^{=}]|\alpha'
\beta'>+<\alpha\beta|[v^{=},\rho_{20}+C^{\perp0}]|\alpha'\beta'>+
        \nonumber\\
& & +\int_{0}^{t}dt'\sum_{\gamma\lambda\gamma'\lambda'}<\alpha\beta|
M^{\gamma\lambda\gamma'\lambda'}(t-t')|\alpha'\beta'>
C^{=}_{\gamma\lambda\gamma'\lambda'}(t'),
\label{3.22}
\eea
with $H^{=}=H^{=}_{0}+v^{=}$,  the memory-kernel
\bea
M^{\gamma\lambda\gamma'\lambda'}(t-t')=1/i \sum_{\mu\nu}N_{\mu}N_{\nu}
K^{\mu\nu}_{\gamma\lambda\gamma'\lambda'}[v^{=},\chi^{\mu}\circ\chi^{\nu*}
e^{-i\Omega_{\mu\nu}(t-t')}],
\label{3.23}
\eea
and the initial correlations
\be
C^{\perp0}=\sum_{\mu\nu}N_{\mu}N_{\nu}C_{\mu\nu}(0)\chi^{\mu}\circ\chi^{\nu*}
e^{-i\Omega_{\mu\nu}t}.
\label{3.24}
\ee
Here, we have used the notations $[A,B]=AB-B^{\dag}A^{\dag}$ and $(A\circ
B)_{1234}=A_{13}B_{24}$.  The coupling with the $ph$-channel produces two
contributions in the $pp$-channel:  (1) A source term originating from
the initial correlations $C_{\mu\nu}(0)$ in the $ph$-channel, and (2) a
retardation term comprising memory effects due to phonon-particle couplings.
The non-locality in time of the memory-kernel -- characterised by a
memory-time $\tau^{*}$ --  is controlled by the number of collective
states contributing to the sum $\sum_{\mu\nu}$ in (\ref{3.23}).
On the other hand, the change in time of $C^{=}(t')$ -- characterised by
a relaxation time $\tau_{rel}$ --  is determined by the strength of the
interaction. Only for $\tau^{*}\ll\tau_{rel}$ one can replace $t'$ by $t$
in $C^{=}(t')$ and extend the integration to infinity (Markov-limit).

The channel-mixing is described by the retardation term in (\ref{3.22}). In
order to allow for a separate treatment of this mixing we introduce a
corresponding mixing correlator $\Delta C$ by $C^{=}=c^{=}+\Delta C$, with
\be
i\dot {c}^{=}=[H^{=},c^{=}]+[v^{=},\rho_{20}+C^{\perp 0}].
\label{3.25}
\ee
Without $C^{\perp0}(t)$ eq. (\ref{3.25}) accounts for a resummation of
ladders as known from time-dependent $G$-matrix theory \cite{9}. The additional
driving term
$[v^{=},C^{\perp0}]$ modifies this by coupling with oscillations due to the
initial correlations in the $ph$-channel. Now, from (\ref{3.22}) we obtain
for the correlations due to a mixing of ladders and loops
\bea
i\dot {\Delta C}=[H^{=},\Delta C]+
\int_{0}^{t}dt'\sum_{\gamma\lambda\gamma'\lambda'}
M^{\gamma\lambda\gamma'\lambda'}(t-t')(\Delta
C(t')+c^{=}(t'))_{\gamma\lambda\gamma'\lambda'}.
\label{3.26}
\eea
In order to study the iterative structure it is more convenient to use the
equivalent integral equation
\bea
\Delta C(t)=\Delta c^{=}(t)+\int^{t}_{0}dt'\sum_{121'2'}W^{121'2'}(t-t')
\Delta C_{121'2'}(t'),
\label{3.27}
\eea
with
\be
\Delta
c^{=}(t)=\int_{0}^{t}dt'\sum_{121'2'}W^{121'2'}(t-t')c^{=}_{121'2'}(t').
\label{3.28}
\ee
The integral kernel $W(t-t')$ may be cast into the form
\be
W^{121'2'}(t-t')=1/i\int^{t-t'}_{0}d\tau
e^{-iH^{=}\tau}M^{121'2'}(t-t'-\tau)e^{iH^{=\dag}\tau} \; ,
\label{3.29}
\ee
which is of first order in the phonon-particle vertices (c.f. (\ref{3.23})),
but of infinite order in the blocked interaction $v^{=}$.
In deriving (\ref{3.28}) and (\ref{3.29}) we have used the identity
$\int^{t}_{0}dt'\int^{t'}_{0}dt''=\int^{t}_{0}dt''\int^{t}_{t''}dt'$.
Applying this relation repeatedly we may arrange the solution of
(\ref{3.27}) in powers of $W$,
\bea
\Delta C_{\alpha\beta\alpha'\beta'}(t)=\Delta
c^{=}_{\alpha\beta\alpha'\beta'}(t)
+\int^{t}_{0}dt'\sum_{121'2'}\left[ W(t-t')\right.
\nonumber\\
+\int^{t}_{t'}d\tau_{1}W(t-\tau_{1})W(\tau_{1}-t')+\int^{t}_{t'}d\tau_{1}\int
^{\tau_{1}}_{t'}d\tau_{2}
\nonumber\\
\left. W(t-\tau_{1})W(\tau_{1}-\tau_{2})W(\tau_{2}-t')
+\cdots \right]^{121'2'}_{\alpha\beta\alpha'\beta'}\Delta c^{=}_{121'2'}(t')
\label{3.30}
\eea
where products of $W$ are formed according to
\be
(WW)^{121'2'}_{\alpha\beta\alpha'\beta'}=\sum_{343'4'}W^{343'4'}_{\alpha\beta
\alpha'\beta'}W^{121'2'}_{343'4'}.
\label{3.31}
\ee
The solution (\ref{3.30}) makes the iterative structure transparent: the
resummation of ladders together with a coupling to vibrations contained in
each $W$ is followed by a resummation of $W$ and thus of phonon-particle
vertices.

Combining all relations with the solution of (\ref{3.25}) we are able
within our approximations to trace back the total two-body correlations $C$ to
the
uncorrelated two-body density $\rho_{20}$ and the initial (vibrational)
correlations $C^{\perp0}$ in the $ph$-channel.
\section{Stationary limit and effective interactions}
In the
stationary limit $\dot{C}=0$ the two-body dynamics can be traced back to
integral equations for frequency-dependent effective interactions. We meet
the stationary limit imposing $\dot{C}^{=}=0$ and $\dot{C}^{\perp}=0$ in
(\ref{3.4}) and (\ref{3.5}), respectively. Strictly speaking, due to the
exchange term
equation (\ref{3.5}) splits up  into two equations for the $ph$-channels
($\alpha,\alpha'$) and ($\alpha,\beta'$). Since, however, the equation for
channel ($\alpha,\beta'$) does not provide new aspects with respect to
channel mixing it is neglected in the following by dropping
${\cal A}_{\alpha\beta}$ in (\ref{3.5}). The price for this simplification is
the loss of antisymmetry with respect to the labels
$(\alpha,\beta)$ and $(\alpha'\beta')$ in the $ph$-channel. Considerations
with three coupled equations are made e.g. in \cite{10} using Green's-function
techniques.

Equations (\ref{3.4}) and (\ref{3.5}) are solved in the stationary limit
by the correlations $C^{=0}$ and $C^{\perp0}$ which can be determined
from the two coupled equations
\bea
(\omega-\epsilon_{\alpha}-\epsilon_{\beta})C^{=0}_{\alpha\beta\alpha'\beta'}
&=&
\langle\alpha\beta|v^{=}(C^{=0}+C^{\perp0}+\rho_{20}^{0})|\alpha'\beta'\rangle
\label{4.1}\\
(\omega-\epsilon_{\alpha}+\epsilon_{\alpha'})C^{\perp0}_{\alpha\beta
\alpha'\beta'}&=&\sum_{\gamma\gamma'}v^{\perp}_{\alpha\gamma'\alpha'\gamma}
(C^{\perp0}+C^{=0})_{\gamma\beta\gamma'\beta'}\;,
\label{4.2}
\eea
where $\rho_{20}^0$ is the stationary uncorrelated two-body density.
This can be seen by subtracting from (\ref{4.1}) and (\ref{4.2}) the
respective hermitean conjugate equation. By introducing  effective
interactions $G(\omega)$ and $\Pi(\omega)$ via the definitions
\bea
\langle\alpha\beta|v(C^{=0}+C^{\perp0}+\rho_{20}^{0})|\alpha'\beta'\rangle
&=&\langle\alpha\beta| G(C^{\perp0}+\rho_{20}^0)|\alpha'\beta'\rangle
\label{4.3}  \\
\sum_{\gamma\gamma'}v^{a}_{\alpha\gamma'\alpha'\gamma}(C^{\perp0}+
C^{=0})_{\gamma\beta\gamma'\beta'}&=&\sum_{\gamma\gamma'}\Pi_{\alpha\gamma'
\alpha'\gamma}C^{=0}_{\gamma\beta\gamma'\beta'}
\label{4.4}
\eea
and insertion of (\ref{4.3}) and (\ref{4.4}) into (\ref{4.1}) and
(\ref{4.2}), respectively, we
obtain  integral equations for the $G$-matrix and the
polarisation matrix $\Pi$ (c.f. also \cite{1})
\bea
G_{\alpha\beta\alpha'\beta'}&=&v_{\alpha\beta\alpha'\beta'}+
\langle\alpha\beta|v(\omega^{(+)}-H^{=}_{0})^{-1}
Q^{=}G|\alpha'\beta'\rangle
\label{4.5} \\
\Pi_{\alpha\beta\alpha'\beta'}&=&v^{a}_{\alpha\beta\alpha'\beta'}+
\sum_{\gamma\gamma'}[v^{a}(\omega^{(+)}-H^{\perp}_{0})^{-1}Q^{\perp}]
_{\alpha\gamma'\alpha'\gamma}\Pi_{\gamma\beta\gamma'\beta'},
\label{4.6}
\eea
with $\omega^{(+)}=\omega + i\eta$, $\eta\to 0^{(+)}$. Assuming again
$\rho^{0}$ and $h(\rho^{0})$ to be diagonal, the matrix elements of the
blocking operators simplify to
$Q^{=}_{\lambda\gamma'\lambda'\gamma}=\delta_{\lambda\lambda'}
\delta_{\gamma\gamma'}(1-n_{\lambda}-n_{\gamma'})$ and $Q^{\perp}_{\lambda
\gamma'\lambda'\gamma}=\delta_{\lambda\gamma}\delta_{\lambda'\gamma'}
(n_{\lambda'}-n_{\lambda})$ and the mean-field two-body propagators
$g^{pp}(\omega)=[\omega^{(+)}-H^{=}_{0}]^{-1}$ and
$g^{ph}(\omega)=[\omega^{(+)}-H^{\perp}_{0}]^{-1}$ reduce to
\bea
g^{pp}_{\alpha\beta\alpha'\beta'}&=&\delta_{\alpha\alpha'}\delta_{\beta\beta'}
[\omega^{(+)}-\epsilon_{\alpha}-\epsilon_{\beta}]^{-1}
\label{4.7}\\
g^{ph}_{\alpha\beta\alpha'\beta'}&=&\delta_{\alpha\beta'}\delta_{\beta\alpha'}
[\omega^{(+)}-\epsilon_{\alpha}+\epsilon_{\alpha'}]^{-1}.
\label{4.8}
\eea
A representation of (\ref{4.5}) and (\ref{4.6}) in terms of  diagrams
is given in Fig. 5 and Fig. 6, respectively. The iterative structure makes
explicit that $G$ accounts for a resummation of ladders while $\Pi$
accounts for a resummation of loops.

Equations (\ref{4.5}) and (\ref{4.6}) describe the stationary limit in
the absence of channel mixing, i.e. when dropping $C^{\perp0}$ in
(\ref{4.1}) and $C^{=0}$ in (\ref{4.2}). In order to account for the
mutual influence of both channels it is tempting to construct a total
$G$-matrix $G^{tot}(\omega)$ in terms of $G(\omega)$
and $\Pi(\omega)$. This can be achieved by the definition
\be
G^{tot}\rho_{20}^{0}=G(C^{\perp0}+\rho_{20}^{0}),
\label{4.9}
\ee
which yields, together with (\ref{4.3}) and (\ref{4.1}), a relation
between $C^{=0}$ and $G^{tot}$. Furthermore, insertion of (\ref{4.4})
into (\ref{4.2}) allows to express $C^{\perp0}$ in terms of $\Pi$
and $C^{=0}$. Finally, combining all relations with (\ref{4.9}) one arrives
at an equation for $G^{tot}$ which may be cast into the form
\bea
G^{tot}_{\alpha\beta\alpha'\beta'}=G_{\alpha\beta\alpha'\beta'}+
\sum_{\lambda\kappa\kappa'}\Gamma
(\omega)_{\alpha\beta\kappa'\alpha'\kappa
\lambda}\frac{1-n_{\kappa}-n_{\lambda}}{\omega^{(+)}-\epsilon_{\kappa}
-\epsilon_{\lambda}}G^{tot}_{\kappa\lambda\kappa'\beta'}.
\label{4.10}
\eea
The frequency-dependent 6-label interaction $\Gamma(\omega)$ is a
combination of $G$, $\Pi$ and a $ph$-propagator,
\be
\Gamma (\omega)_{\alpha\beta\kappa'\alpha'\kappa\lambda}=
\sum_{\gamma}G_{\alpha\beta\gamma \lambda}\frac{n_{\alpha'}-n_{\gamma}}
{\omega^{(+)}-\epsilon_{\gamma} +\epsilon_{\alpha'}}\Pi_{\gamma\kappa'
\alpha'\kappa}.
\label{4.11}
\ee
Eq. (\ref{4.10}) is not an usual Bethe-Goldstone equation since not only two
but three labels of $G^{tot}$ are involved in the summation on the $r.h.s.$.
Its iterative structure is displayed in Fig. 7. The successive iteration
proceeds in powers of $\Gamma$ connected by a respective $pp$-propagator.
A diagram of second order in $\Gamma$ is shown in Fig. 8.

The effective interaction (4.11) emerging from the limit NQCD in the
stationary limit accounts for the lowest order parquet diagrams as pointed
out in \cite{10}. In general, the parquet formalism \cite{P1,P2,P4}
adresses the problem of summing up irreducible interaction diagrams
(including the mixed channels) with proper symmetry (for bosons) or
antisymmetry (for fermions) without double counting of interactions. It thus
includes potentially much higher classes of diagrams than those considered
in (4.11) due to the neglect of $C_3$ and higher order correlations.
Within the notation
of \cite{P4} our effective interaction $\Gamma(\omega)$ corresponds
to the parquet diagrams including the $s, t$ and $u$ channels of the
bare interaction, however, without "left-" and "right-hand" vertex
corrections. It is presently still a matter of debate if the higher-order
diagrams -- that potentially are included in the parquet formalism -- are
actually needed for a "proper" description of two-body dynamics in the
nuclear physics context. We note in passing that the parquet theory also
has been formulated on the three-body level in \cite{P4} while a
corresponding correlation dynamics of Green's functions up to the
three-body level has been presented in \cite{WZC}.

Equation (\ref{4.10}) describes nonperturbatively the channel mixing in the
stationary limit of two-body dynamics. For transparency $\rho^{0}$ was
assumed to be diagonal. This assumption can easily be dropped replacing
$1-n_{\kappa}-n_{\lambda}$ in (\ref{4.10}) and $n_{\alpha'}-n_{\gamma}$
in (\ref{4.11}) by the non-diagonal expressions for $Q^{=}$ and $Q^{\perp}$,
respectively. For a complete description of a stationary system the equations
for the effective interactions must be completed by the stationary
one-body equation
\be
0=[\hat t,\rho^{0}]+tr_{2}[v,\rho^{0}_{20}+C^0].
\label{4.12}
\ee
Here, $\hat t$ is the kinetic energy operator, and $tr_{2}$ means the
trace over the second particle. With $C^{0}=C^{\perp0}+C^{=0}$ and the
definitions (\ref{4.3}) and (\ref{4.9}) we obtain
\be
0=[\hat
t,\rho^{0}]+tr_{2}[G^{tot},\rho^{0}_{20}].
\label{4.13}
\ee
This equation for
$\rho^{0}$ together with the equations for $G^{tot}$, $G$ and $\Pi$
represent a closed set of equations for the stationary limit of the system
which includes channel mixing and therefore accounts
nonperturbatively for the interplay of long- and short-range
correlations.
\section{In-medium scattering approach}
Short-range correlations are associated with multiple $pp$- (or $hh$-)
collisions mediated by the in-medium interaction $v^{=}$. The in-medium
scattering approach is based on the assumption that these collisions happen
on a time-scale which is short as compared with the time in between the
collisions. Neglecting, for a  moment, long-range correlations and
approximating the ``collision-free'' evolution on the long time-scale by
the uncorrelated two-body density,
\be
i\dot{\rho}_{20}=[H^{=}_{0},\rho_{20}],
\label{5.1}
\ee
we obtain for the short-range correlations \cite{7,9}
\be
C^{=}=-\rho_{20}+\Omega(E)\rho_{20}\Omega(E)^{\dag},
\label{5.2}
\ee
where the two-body Moeller operator $\Omega(E)$ depends on the total energy
of the ``scattering-system'' and follows the equation
\be
\Omega(E)=1+[E^{+}-H^{=}_{0}]^{-1}v^{=}\Omega(E).
\label{5.3}
\ee
Equation (\ref{5.3}) is equivalent to (\ref{4.5}) when defining
the $G$-matrix by $G(E)=v\Omega(E)$.

Within this concept of two different time-scales long-range correlations
are included assuming that they are operative essentially on the long
time-scale, i.e. in between the collisions. This allows to apply the same
integration procedure from \cite{3,7,9} to eq. (\ref{3.2}) which yields
\be
C^{=}\approx -\rho_{20}-C^{\perp}+\Omega(\rho_{20}+C^{\perp})\Omega^{\dag}.
\label{5.4}
\ee
The Moeller-operators may be expressed in terms of G as
\be
\Omega(E)=1+g(E)Q^{=}G(E)
\label{5.5}
\ee
with $g(E)=[E^{+}-H^{=}_{0}]^{-1}$. Insertion of (\ref{5.4}) into (\ref{3.8})
provides an equation for the long-range correlations
\bea
i\dot{C}^{\perp}&=&(1+{\cal P}{\cal P})H^{\perp}C^{\perp}+(1+{\cal P}{\cal
P})v^{\perp}([gQ^{=}G,C^{\perp}]_{+}+  \nonumber \\
 & & gQ^{=}GC^{\perp}G^{\dag}Q^{=}g^{\dag})+D(\rho_{20}).
\label{5.6}
\eea
We observe that channel mixing produces a driving term
\be
D(\rho_{20})=(1+{\cal P}{\cal P})
v^{\perp}(\Omega\rho_{20}\Omega^{\dag}-\rho_{20})
\label{5.7}
\ee
as well as an  additional interaction term of mixed structure (second term on
the r.h.s.): $C^{\perp}$ is horizontally
connected with G but the resulting expression is vertically connected with
$v^{\perp}$. Both terms are non-hermitean due to the propagator $g(E)$.

Equation (\ref{5.6}) holds for an arbitrary single-particle basis.
Together with (\ref{2.3}) and (\ref{5.3}) it represents a closed set of
equations for the long-range correlations $C^{\perp}$, the $G$-matrix $G$
and the one-body density $\rho$.
The one-body equation reads, with (\ref{3.1}) and (\ref{5.4}),
\bea
i\dot{\rho}-[\tilde h,\rho]=I_{coll}(\rho)+tr_{2}[G,C^{\perp}]+
 tr_{2}[GC^{\perp}G^{\dag},Q^{=}g^{\dag}],
\label{5.8}
\eea
with the on-shell collision term from time-dependent $G$-matrix theory
\bea
I_{coll}=i \; tr_{2}([ImG,\rho_{20}]_{+}-[G\rho_{20}G^{\dag},Q^{=}Img]_{+})
\label{5.9}
\eea
and the renormalised mean field $\tilde h=\hat t+tr(ReG^{a}\rho)$. Equation
(\ref{5.8}) was used in \cite{11} to study nuclear damping without
channel mixing, i.e. with approximate long-range correlations which follow
from (\ref{3.8}) when dropping the term with $C^{=}$.

Due to the non-hermitean character of (\ref{5.6}) we expect that the mixing
with the  $pp$-channel leads to damped vibrations in the $ph$-channel.
To make this more transparent we assume - as in section 3 - that the
one-body density appearing in the interaction terms of (\ref{5.6}) is
diagonal and time-independent.
This allows for a transformation into the RPA-basis
defined in (\ref{3.10}, \ref{3.11}). The result is
\be
i\dot{C}_{\mu\nu}=\Omega_{\mu\nu}C_{\mu\nu}+\sum_{\mu'\nu'}K^{\mu\nu}_{\mu'\nu'}
C_{\mu'\nu'}+D_{\mu\nu}(\rho_{20})
\label{5.10}
\ee
with the complex coupling matrix
\bea
K^{\mu\nu}_{\mu'\nu'}&=&N_{\mu'}N_{\nu'}\sum_{1234}K^{\mu\nu}_{1234}
T^{\mu'\nu'}_{1234}
\label{5.11}   \\
T^{\mu'\nu'}_{1234}&=&
\langle12|(\Omega\chi^{\mu'}\circ\chi^{\nu'\ast}\Omega^{\dag}
-\chi^{\mu'}\circ\chi^{\nu'\ast})|34\rangle
\label{5.12}
\eea
and the driving term
\be
D_{\mu\nu}(\rho_{20})=\sum_{1234}K^{\mu\nu}_{1234}
\langle12|(\Omega\rho_{20}\Omega^{\dag}-\rho_{20})|34\rangle.
\label{5.13}
\ee
The contributions to the sums in (\ref{5.11}) and (\ref{5.13}) factorize
into a term $K^{\mu\nu}_{1234}$ which is, according to (\ref{3.18}), a linear
combination of phonon-particle vertices, multiplied by a term which comprises
the short-range correlations in form of Moeller operators $\Omega(E)$.
In the limit $\Omega=1$, i.e. without collisions, both coupling matrix
and driving term vanish. The source for damping is the imaginary part of
the coupling matrix. Using (\ref{5.5}) we obtain
\bea
ImK^{\mu'\nu'}_{\mu\nu}&=&-N_{\mu'}N_{\nu'}\sum_{1234}K^{\mu\nu}_{1234}
\langle 12|(1+R)\chi^{\mu'}\circ\chi^{\nu'\ast}S^{\dag}-
                   \nonumber\\
 & & -S\chi^{\mu'}\circ\chi^{\nu'\ast}(1+R^{\dag})|34 \rangle,
\label{5.14}
\eea
with
\bea
R\approx -ImgQ^{=}ImG
\label{5.15}
\eea
\bea
S\approx ImgQ^{=}ReG.
\label{5.16}
\eea
Here we have adopted the on-shell approximation
\be
g(E)\approx iImg(E)=-i\pi \delta(E-h(1)-h(2)).
\ee
A further evaluation of (\ref{5.10}) is achieved by solving the
non-hermitean eigenvalue-problem
\bea
\omega_{\eta}\xi^{\eta}=A\xi^{\eta}
\label{5.17} \\
A^{\dag}\psi^{\eta\dag}=\omega^{\ast}_{\eta}\psi^{\eta\dag}
\label{5.18}
\eea
with $A_{\mu\nu\mu'\nu'}=(\Omega_{\mu}-\Omega_{\nu})\delta_{\mu\mu'}\delta_
{\nu\nu'}+K^{\mu'\nu'}_{\mu\nu}$. The eigenstates $\xi^{\eta}_{\mu\nu},\psi^
{\eta\ast}_{\mu\nu}$  form a biorthogonal system which follows the
orthogonality
relation
\be
\sum_{\mu\nu}\psi^{\eta\ast}_{\mu\nu}\xi^{\eta'}_{\mu\nu}=\delta_{\eta\eta'}.
\label{5.19}
\ee
An expansion $C_{\mu\nu}(t)=\sum_{\eta}C_{\eta}(t)\xi^{\eta}_{\mu\nu}$ yields,
together with (\ref{5.10}), an equation for the components $C_{\eta}(t)$,
the solution of which is given by
\bea
C_{\eta}(t)=e^{-i\omega_{\eta}t}C_{\eta}(0)+1/i\int^{t}_{0}dt'e^{-i\omega_{\eta}
(t-t')}\sum_{\mu\nu}\xi^{\eta\dag}_{\mu\nu}D_{\mu\nu}(\rho_{20}(t')).
\label{5.20}
\eea
A transformation back into the single-particle basis provides for the
long-range
correlations
\bea
C^{\perp}_{\alpha\beta\alpha'\beta'}& \approx &\sum_{\eta}f^{\eta}_{\alpha\beta
\alpha'\beta'}e^{-i\omega_{\eta}t}C_{\eta}(0)+
\nonumber\\
 & & 1/i\int_{0}^{t}dt'\sum_{\eta,1234}
f^{\eta}_{\alpha\beta\alpha'\beta'}K^{\eta}
_{1234}e^{-i\omega_{\eta}(t-t')}(\Omega\rho_{20}(t')\Omega^{\dag}-\rho_{20}(t'))
_{1234}  \nonumber\\
\label{5.21}
\eea
with
\bea
f^{\eta}_{1234}&=&\sum_{\mu\nu}N_{\mu}N_{\nu}\chi^{\mu}_{13}\chi^{\nu\ast}_{24}
\xi^{\eta}_{\mu\nu}
\label{5.22}\\
K^{\eta}_{1234}&=&\sum_{\mu\nu}\psi^{\eta\ast}_{\mu\nu}K^{\mu\nu}_{1234}.
\label{5.23}
\eea
Expression (\ref{5.21}) is our central result. The long-range correlations
are superpositions of vibrational modes which are damped due to the coupling
with short-range correlations. This collisional damping is
non-perturbative (i) with respect to the phonon-particle vertices due to
the diagonalisation of $A$ (\ref{5.17}, \ref{5.18}) and (ii) with respect to
the
two-body collisions due to the $G$-matrix appearing in (\ref{5.14}).
Furthermore, after insertion of (\ref{5.21}) into (\ref{5.8}), we observe
that long-range correlations produce vibrational terms
in the one-body equation.

\section{Summary}
In this paper we have presented a reformulation of two-body dynamics in which
the equation for the two-body correlation function -- as obtained from the
time-dependent density matrix hierarchy -- is replaced by two coupled
equations for correlations in the $pp$- and the $ph$-channel.
These correlations are addressed as short- and long-range correlations
and associated with in-medium collisions and vibrational motion, respectively.
Their mutual influence (termed channel-mixing) is analysed by means of an
integral equation for the corresponding mixing correlations which exhibits
the iterative structure of two-body dynamics: After a resummation of ladders
accounting for in-medium collisions a resummation of loops in terms of
vibrational RPA-modes is performed which accounts for the particle-phonon
coupling. In this way the complete two-body correlation function is traced
back to the uncorrelated two-body density and the initial vibrations in the
$ph$-channel. We stress that our approach is
non-perturbative also with respect to mixed diagrams and that
double-counting of interactions is avoided.

In the stationary limit the channel-mixing can be traced back to an effective
interaction which combines the usual $G$-matrix with the polarisation matrix.
The resulting equations may be viewed as basis for a more sophisticated
treatment of the true (correlated) ground state in the sense that channel
mixing and hence the mutual influence of long- and short-range correlations
is nonperturbatively included.

Assuming that short- and long-range correlations evolve on two different
time-scales the short-range correlations can be integrated in terms of the
$G$-matrix. In this in-medium scattering approach (termed time-dependent
$G$-matrix theory when neglecting long-range correlations) channel-mixing
shows up in the damping of the vibrational modes which constitute the
long-range correlations. In the one-body equation in-medium collisions care
for both a renormalisation of the mean field and a collision term, whereas
long-range correlations produce collective effects on the one-body level in
terms of vibrations which are damped due to channel mixing.

The importance of channel-mixing has been, furthermore, demonstrated
by numerical results for the
total energy which approaches the true ground state energy much closer than
without this mixing.
In addition, the relative importance of higher order correlations is found
to decrease substantially with increasing order of the
correlations thus indicating the convergence of the cluster expansion
at least for configurations close to the ground state, i.e. for
temperatures up to a few MeV.

\newpage

\section*{Appendix A}

The cluster expansion for the density matrices $\rho_n$, that is needed for
the derivation of (2.1) and (2.2) as well as for the trace relations
(2.6) - (2.8) up to the four-body level is given by

$$
\rho _1(11') = \rho (11') ,
$$
\bfr                   (A1)
\efr
$$
\rho _2(12,1'2') = {\cal A}_{1'2'}\rho (11')\rho (22') + C_2(12,1'2') =
\rho _{20}(12,1'2') + C_2(12,1'2'),
$$
\bfr      (A2)
\efr
$$
\rho _3(123,1'2'3') =
{\cal S}_{1'}{\cal A}_{2'3'}\rho (11')\rho (22')\rho (33') +
{\cal S}_{1'}\rho (11')C_2(23,2'3')
$$
$$
+ {\cal S}_{2'}\rho (22')C_2(13,1'3') + {\cal S}_{3'}\rho (33')C_2(12,1'2') +
C_3(123,1'2'3'),
$$
\bfr  (A3)
 \efr
$$
\rho _4(1234,1'2'3'4') =
{\cal A}_{1'2'}{\cal S}_{3'}\Lambda _{4'}\rho (11')\rho (22')\rho (33')\rho (%
44')
$$
$$
+ \Lambda _{1'}(1-{\cal P}_{2'3'}-{\cal P}_{2'4'}) \rho (11')\rho
(22')C_2(34,3'4')
$$
$$
+ \Lambda _{1'}(1-{\cal P}_{2'4'}-{\cal P}_{3'4'}) \rho (11')\rho
(44')C_2(23,2'3')
$$
$$
+ \Lambda _{2'}(1-{\cal P}_{1'3'}-{\cal P}_{3'4'}) \rho (22')\rho
(33')C_2(14,1'4')
$$
$$
+ \Lambda _{2'}(1-{\cal P}_{1'4'}-{\cal P}_{3'4'}) \rho (22')\rho
(44')C_2(13,1'3')
$$
$$
+ \Lambda _{3'}(1-{\cal P}_{1'4'}-{\cal P}_{2'4'}) \rho (33')\rho
(44')C_2(12,1'2')
$$
$$
+ \Lambda _{3'}(1-{\cal P}_{2'3'}-{\cal P}_{3'4'}) \rho (33')\rho
(11')C_2(24,2'4')
$$
$$
+ \Gamma _{2'}C_2(12,1'2')C_2(34,3'4') + \Gamma _{3'}C_2(13,1'3')C_2(24,2'4')
+ \Gamma _{4'}C_2(14,1'4')C_2(23,2'3')
$$
$$
+ \Lambda _{1'}\rho (11')C_3(234,2'3'4') +
\Lambda _{2'}\rho (22')C_3(134,1'3'4')
$$
$$
+ \Lambda _{3'}\rho (33')C_3(124,1'2'4') +
\Lambda _{4'}\rho (44')C_3(123,1'2'3')
$$
$$
+ C_4(1234,1'2'3'4'),
$$
\bfr      (A.4)
\efr

\noindent
with the two- and three-body antisymmetrization operators
$$
{\cal A}_{ij} = 1 - {\cal P}_{ij} ; \   {\cal A}_{i'j'} = 1 - {\cal P}_{i'j'} ,
$$
\bfr  (A.5)
\efr
$$
{\cal S}_{1'} = 1-{\cal P}_{1'2'}-{\cal P}_{1'3'} ;
\  {\cal S}_1 = 1-{\cal P}_{12}-{\cal P}_{13} ,
$$
$$
{\cal S}_{2'} = 1-{\cal P}_{1'2'}-{\cal P}_{2'3'} ;
\   {\cal S}_2 = 1-{\cal P}_{12}-{\cal P}_{23},
$$
$$
{\cal S}_{3'} = 1-{\cal P}_{1'3'}-{\cal P}_{2'3'} ;
\   {\cal S}_3 = 1-{\cal P}_{13}-{\cal P}_{23}.
$$
\bfr  (A6)
\efr
The four-body antisymmetrization operators in (A4), furthermore, read
explicitly
$$
\Gamma _{2'} = (1-{\cal P}_{1'3'}-{\cal P}_{1'4'}-{\cal P}_{2'3'}-
{\cal P}_{2'4'}+{\cal P}_{1'3'}{\cal P}_{2'4'}),
$$
$$
\Gamma _2= (1-{\cal P}_{13}-{\cal P}_{14}-{\cal P}_{23}-
{\cal P}_{24}+{\cal P}_{13}{\cal P}_{24}),
$$
$$
\Gamma _{3'} = (1-{\cal P}_{1'2'}-{\cal P}_{1'4'}-
{\cal P}_{2'3'}-{\cal P}_{3'4'}+{\cal P}_{1'2'}{\cal P}_{3'4'}),
$$
$$
\Gamma _3= (1-{\cal P}_{12}-{\cal P}_{14}-{\cal P}_{23}-
{\cal P}_{34}+{\cal P}_{12}{\cal P}_{34}),
$$
$$
\Gamma _{4'} = (1-{\cal P}_{1'2'}-{\cal P}_{1'3'}-
{\cal P}_{2'4'}-{\cal P}_{3'4'}+{\cal P}_{1'3'}{\cal P}_{2'4'}),
$$
$$
\Gamma _4= (1-{\cal P}_{12}-{\cal P}_{13}-{\cal P}_{24}-
{\cal P}_{34}+{\cal P}_{13}{\cal P}_{24}),
$$
$$
\Lambda _{1'} = (1-{\cal P}_{1'2'}-{\cal P}_{1'3'}-{\cal P}_{1'4'});
\   \Lambda _1 = (1-{\cal P}_{12}-{\cal P}_{13}-{\cal P}_{14}),
$$
$$
\Lambda _{2'} =  (1-{\cal P}_{2'1'}-{\cal P}_{2'3'}-{\cal P}_{2'4'});
\   \Lambda _2 = (1-{\cal P}_{21}-{\cal P}_{23}-{\cal P}_{24}),
$$
$$
\Lambda _{3'} = (1-{\cal P}_{3'1'}-{\cal P}_{3'2'}-{\cal P}_{3'4'});
\    \Lambda _3 = (1-{\cal P}_{31}-{\cal P}_{32}-{\cal P}_{34}),
$$
$$
\Lambda _{4'} = (1-{\cal P}_{4'1'}-{\cal P}_{4'2'}-{\cal P}_{4'3'});
\    \Lambda _4 = (1-{\cal P}_{41}-{\cal P}_{42}-{\cal P}_{43}),
$$
\bfr  (A.7)
\efr
\noindent
while ${\cal P}_{ij}$ denotes the exchange operator between particle $i$ and
$j$.
\newpage
\section*{Figure captions}
Fig. 1: Time average of the total traces
$tr \ C_2$, $tr \ C_3$ and $tr \ C_4$ as a
function of the initialization temperature T within the trace-conserving
SCD-approach for a model $^{16}O$-nucleus (c.f. \cite{1}).
The time-average was performed over a time period
of $80\times 10^{-23}s$ after the in-medium interactions have been
switched on adiabatically with $t_s=100 \times 10^{-23}s$ (see section 2.2).
For a better comparison $tr \ C_3$ and $tr \ C_4$ are multiplied with constant
factors 10 and 100, respectively.
\vspace{0.9cm}\\
Fig. 2: Total energy of the model $^{16}O$-nucleus -- initialized with
$T=0$ MeV temperature -- as a function
of the switch-on time $t_s$ in the limits BORN, RPA, TDGMT, NQCD and SCD
(see text).
\vspace{0.9cm}\\
Fig. 3: Monopole response of the $^{16}O$ model nucleus in the
occupation numbers.
The calculation was performed within the NQCD approximation with
$T=1$ MeV initialization temperature.
The $^{16}O$-nucleus was excited by a $15$ MeV monopole boost
at $t_b = 120 \times 10^{-23}s$ after switching on the in-medium
interaction with $t_s=100\times 10^{-23}s$.
\vspace{0.9cm}\\
Fig. 4: Monopole response in the two-body correlation matrix elements.
The calculation is the same as in Fig. 3.
The four diplayed curves correspond to
$\beta =$ 1 (solid line), 4 (dashed line), 5 (dashed-dotted line)
and 7 (dotted line).
\vspace{0.9cm}\\
Fig. 5: Integral equation for the $G$-matrix. \vspace{0.9cm}\\
Fig. 6: Integral equation for the polarisation matrix. \vspace{0.9cm}\\
Fig. 7: Integral equation for $G^{tot}$; the operators $G$ and $\Pi$
in the second diagram on the r.h.s. are connected by one line forming
the 6-label operator $\Gamma$. \vspace{0.9cm}\\
Fig. 8: Second-order contribution in $\Gamma$.
\end{document}